\documentclass[a4paper,twoside]{article}
\usepackage[utf8]{inputenc}
\usepackage{epsfig}
\usepackage{subcaption}
\usepackage{calc}
\usepackage{amssymb}
\usepackage{amstext}
\usepackage{amsmath}
\usepackage{amsthm}
\usepackage{multicol}
\usepackage{pslatex}

\usepackage{natbib}
\let\bibhang\relax
\usepackage{apalike}

\usepackage[nolist]{acronym} 
\usepackage{hyperref} 

\usepackage{enumitem}
\usepackage{setspace}

\usepackage{tikz}
\usetikzlibrary{arrows,chains,positioning,fit, shapes,calc,backgrounds,shapes.multipart}

\usepackage{xltabular}
\usepackage{lipsum}
\usepackage{booktabs}
\usepackage[draft]{tikzpeople}

\usepackage{xcolor}
\usepackage{tabu}
\usepackage[shortcuts]{extdash}
\usepackage{SCITEPRESS}     

\begin{document}
\begin{acronym}
\acro{cc}[CC]{Control Center}
\acroplural{cc}[CCs]{Control Centers}

\acro{cr}[CR]{Control Room}
\acroplural{cr}[CR]{Control Rooms}

\acro{ddt}[DDT]{Dynamic Driving Tasks}
\acroplural{ddt}[DDTs]{Dynamic Driving Tasks}
\acro{oedr}[OEDR]{Object and Event Detection and Response}

\acro{av}[AV]{Automated Vehicle}
\acroplural{av}[AVs]{Automated Vehicles}

\acro{avf}[AVF]{Automated Vehicle Fleet}
\acroplural{avf}[AVF]{Automated Vehicle Fleets}

\acro{ad}[AD]{Automated Driving}

\acro{ad}[AD]{Automated Driving}

\acro{ads}[ADS]{Automated Driving System}
\acroplural{ads}[ADS]{Automated Driving Systems}

\acro{odd}[ODD]{Operational Design Domain}

\acro{rd}[RD]{Remote Driver}
\acroplural{rd}[RD]{Remote Drivers}

\acro{soc}[SOC]{State of Charge}

\acro{ca}[CA]{Control Action}
\acroplural{ca}[CAs]{Control Actions}

\acro{cf}[CF]{Causal Factor}
\acroplural{cf}[CFs]{Causal Factors}

\acro{uca}[UCA]{Unsafe Control Action}
\acroplural{uca}[UCAs]{Unsafe Control Actions}

\acro{stpa}[STPA]{System-Theoretic Process Analysis}

\acro{hmd}[HMD]{Head Mounted Display}
\acroplural{hmd}[HMDs]{Head Mounted Displays}

\acro{oem}[OEM]{Original Equipment Manufacturer}
\acroplural{oem}[OEMs]{Original Equipment Manufacturer}

\acro{cbw}[CbW]{Conduct by Wire}

\acro{adas}[ADAS]{Advanced Driver Assistance Systems}

\acro{hmi}[HMI]{Human Machine Interface}

\acro{oem}[OEM]{Original Equipment Manufacturer}
\acro{ul}[UL]{up link}
\acro{dl}[DL]{down link}

\acro{swa}[SWA]{Steering Wheel Angle}
\acro{sc}[SC]{System-level Constraint}
\acroplural{sc}[SCs]{System-level Constraints}

\acro{hazop}[HAZOP]{Hazard and Operability Study}
\acro{fta}[FTA]{Fault Tree Analysis}
\acro{fmea}[FMEA]{Failure Mode and Effect Analysis}
\acro{stamp}[STAMP]{Systems-Theoretic Accident Model and Processes}
\acro{sotif}[SOTIF]{Safety of the Intended Functionality}

\acro{swOp}[SW]{steering wheel actuator}
\acro{bpOp}[BP]{brake pedal actuator}
\acro{tpOp}[TP]{throttle pedal actuator}
\acro{imu}[IMU]{Inertial Measurement Unit}

\acro{swaOp}[$\dot{\delta}_{\text{H,O}}$]{change steering wheel command}
\acro{sBp}[$s_{\text{Bp}}$]{brake pedal travel}
\acro{sTp}[$s_{\text{Tp}}$]{throttle pedal travel}
\acro{vD}[$v_{\text{d}}$]{desired velocity}
\acro{vA}[$v_{\text{a}}$]{actual velocity}
\acro{swaD}[$\delta_{\text{d}}$]{desired steering wheel angle}
\acro{swaA}[$\delta_{\text{a}}$]{actual steering wheel angle}
\acro{swaM}[$M_{\text{H}}$]{steering torque}
\acro{engineM}[$M_{\text{E}}$]{engine torque}

\end{acronym}
\title{Systems-Theoretic Safety Assessment of Teleoperated Road Vehicles}

\author{\authorname{Simon Hoffmann\sup{1}, Dr. Frank Diermeyer\sup{1}}
\affiliation{\sup{1}Institute of Automotive Technology, Technical University of Munich, Boltzmannstr. 15, Garching b. M\"unchen, Germany}
\email{\{hoffmann, diermeyer\}@ftm.mw.tum.de}
}


\abstract{
Teleoperation is becoming an essential feature in automated vehicle concepts, as it will	 help the industry overcome challenges facing automated vehicles today. Teleoperation follows the idea to get humans back into the loop for certain rare situations the automated vehicle cannot resolve. Teleoperation therefore has the potential to expand the operational design domain and increase the availability of automated vehicles. This is especially relevant for concepts with no backup driver inside the vehicle.
While teleoperation resolves certain issues an automated vehicle will face, it introduces new challenges in terms of safety requirements. While safety and regulatory approval is a major research topic in the area of automated vehicles, it is rarely discussed in the context of teleoperated road vehicles. The focus of this paper is to systematically analyze the potential hazards of teleoperation systems. An appropriate hazard analysis method (\acs{stpa}) is chosen from literature and applied to the system at hand. The hazard analysis is an essential part in developing a safety concept (e.g., according to ISO26262) and thus far has not been discussed for teleoperated road vehicles.
}
\onecolumn \maketitle \normalsize \setcounter{footnote}{0} \vfill

\section{\uppercase{Introduction}}
\label{sec:introduction}
\noindent \ac{ad} is, besides electrification of road vehicles, one of the biggest challenges and opportunities the automotive industry is currently facing. \acp{oem}, as well as research institutes, are investing significant effort into getting \acp{ads} \footnote{Taxonomy according to \cite{SAE2018a} is used throughout this paper} on public roads.
To describe the degree of automation of a specific driving automation system, a taxonomy was introduced \citep{SAE2018a}. This taxonomy differentiates between six different levels from L0 ``No Driving Automation'' to L5 ``Full Driving Automation.'' Only L3 - L5 systems, which are capable of performing the \ac{ddt} on a ``sustained basis'' \citep{SAE2018a}, are considered \acp{ads}. Below L3, even if the vehicle is performing longitudinal and lateral driving tasks on a sustained basis, the human driver is responsible for the \ac{oedr}. For L3-\ac{ads}, the system performs the whole \ac{ddt} including \ac{oedr}. If a \ac{ddt} performance-relevant system failure occurs or when the driving automation system is about to leave its \ac{odd}, the fallback-ready user has to take over. Up to this point, a human driver is required inside the vehicle. As indicated by \citet{Abe2019}, applications such as public transportation or taxi services could strongly benefit if human drivers are replaced by the \ac{ads} and the system is responsible for \ac{ddt}-fallback itself (L4+). According to \citet{SAE2018a}, the system has to perform a \ac{ddt}-fallback by achieving a minimal risk condition, which could be a safe stop at an appropriate place. This, however, requires the \ac{ads} to be fully functional and has to be separated from the failure mitigation strategy, which is required to stop the vehicle in case of critical system failures.\\
The previous paragraph shows the different levels of \ac{ads} and the role of humans in context of the driving task. \cite{SAE2018a} indicates that even an L4+ vehicle, which does not require a user inside the vehicle, depends on some fallback strategies to stop the vehicle in certain situations. A fallback driver might still be able to perform the \ac{ddt} and might not be dependent on degraded or failing system components. The absence of such a fallback driver results in the vehicle and its passengers being stranded and obstructing traffic. The higher the reliability of such a system, the higher the acceptance and profitability of \ac{ads} will be. \ac{ads} are getting better over time. However, taking into account every edge case appearing on public roads might not be feasible. Therefore, having a reliable fallback option for L4+ vehicles will not only decrease the time-to-market launch but also the acceptance and profitability of L4+ \ac{ads} e.g., in public transportation or logistics.\\
To solve this problem, the concept of teleoperation can be used \citep{Georg2018, Bout2017}. After the \ac{ads} dedicated vehicle comes to a stop by way of its integrated \ac{ddt}-fallback function or following a failure mitigation strategy, the vehicle contacts a control center. A concept for such a control center is proposed by \cite{Feiler2020}. A remote operator has to analyze the problem and can choose among different options to resolve the situation \citep{Feiler2020}. One of them remotely controls the vehicle until the \ac{ads} can continue the \ac{ddt} itself. Alternatively, the operator could teleoperate the vehicle to the next bus station or to a vehicle repair center.\\
Vehicle sensor information is sent to the operator via the cellular network. The operator has to comprehend the situation and the vehicle's surroundings based on the sensor information and send control signals back to the vehicle. This introduces new problems to the system, such as latencies, reduced situation awareness or connection losses. The presented work analyzes those problems and identifies further hazards related to teleoperated road vehicles. This is a necessary step in developing a holistic safety concept for teleoperated road vehicles. Before teleoperated road vehicles are analyzed in \autoref{sec:stpaAppl}, a short overview on related work regarding safety assessment and teleoperation is given in \autoref{sec:relatedwork}. Further a short overview on the used method (\autoref{sec:stpa}), and the system this method is applied to (\autoref{sec:systemDescription}), is provided.

\vspace{-0.2cm}
\section{\uppercase{Related Work}}
\label{sec:relatedwork}
\noindent Before analyzing the system of teleoperated road vehicles, a literature review on their problems and also identified solutions is given. Furthermore the advantages of a systems-theoretic approach are outlined and related work regarding its application in a automotive context is presented.
\subsection{Teleoperation of Road Vehicles}
\label{sec:relWorkTof}
Teleoperation is a widely used concept for different applications. It is often utilized to reach hazardous or inaccessible areas, such as in space-robotics or deep-sea exploration. \citet{Bensoussan1997} apply this concept to road vehicles for distributing car sharing vehicles. With the development of \ac{ad}, the focus of teleoperation is to provide a fallback for \ac{ads}. However, teleoperation itself is prone to some problems that are the focus of research as long as this research area exists.\\
\citet{Adams1961} shows the decreasing performance of humans in path-following experiments depending on transmission latencies. \citet{Sheridan1963} and \citet{Ferrell1965} show increasing task-completion times with increasing delay. Variable latencies in the context of driving tasks are investigated by \citet{Davis2010}, \citet{Gnatzig2013a} and \citet{Liu2017}. Variable delay are shown to be even worse than constant transmission delay for driving tasks. Different solutions are proposed to overcome the negative impact of latencies for teleoperated road vehicles. \citet{Chucholowski2016} proposed a predictive display, which increases driving performance under delay. \citet{Gnatzig2012}, \citet{Hosseini2014a} and \citet{Fong2001a} utilize more automation on the robot side by sending trajectories or waypoints to the vehicle. Certain control loops are closed within the robot and the operator does not act on a stabilization level, which is prone to latencies. \citet{Lichiardopol2007} provides a categorization of the different teleoperation concepts and the human and robot responsibilities. \citet{Tang2014e} propose a method that takes into account connection losses in teleoperated road vehicles and brings the vehicle to a safe stop.\\
Another teleoperation problem is the situation awareness of the operator not being located in the vehicle. \citet{Georg2018}, \citet{Hosseini2016} and \citet{Bout2017} investigate the influence of \acp{hmd} e.g., on situation awareness. \citet{Georg2020} investigates the effects of videoquality, videocanvases and displays on situation awareness. \citet{Hosseini2016a} and \citet{Schimpe2020} propose solutions to support the operator with the driving task and overcome the negative effects of situation awareness and latencies regarding collisions.

\subsection{Safety Assessment}
Section \ref{sec:relWorkTof} shows that different aspects reducing the safety of teleoperated road vehicles are already addressed in research. Additionally, different concepts and solutions are proposed in literature to overcome certain problems. To ensure functional safety in an automotive context, the ISO~26262 standard exists \citep{ISO.2018}. Section 3 of ISO~26262, which results in a functional safety concept, requires a hazard and risk analysis of the system at hand. Usually, hazard analysis methods such as \ac{hazop}, \ac{fmea} or \ac{fta} are applied to systematically identify potential hazards of the system. 
According to \citet{Placke2015}, traditional methods tend to focus on component failures. However, accidents often happen due to component interaction, regardless of individual components or software working correctly \citep{Thomas2015a}. Software-related accidents are often caused by flawed requirements instead of coding errors. However, flawed requirements are hard to capture using traditional failure-based methods \citep{Thomas2015a}.\\
\noindent To complement functional safety covered by ISO~26262, ISO/PAS~21448 addresses the \ac{sotif} \citep{ISO.2019}. The scope of ISO/PAS~21448 is to address hazards, resulting from functional insufficiencies of the intended functionality or foreseeable misuse \citep{ISO.2019}. Besides the previously mentioned hazard analysis methods, \ac{stpa} is listed in ISO/PAS~21448.
\ac{stpa} was proposed by \citet{Leveson2011}, to overcome certain flaws in the existing methods. This system engineering approach follows the idea to formulate safety as a control problem rather than a reliability problem \citep{Leveson2011}. \ac{stpa} has the ability to consider interactions between different types of components, such as software, hardware or humans \citep{Placke2015}. According to \citet{Thomas2015a}, ``STPA is a top-down hazard analysis method designed to go beyond traditional component failures to also identify problems such as dysfunctional interactions, flawed requirements, design errors, external disturbances, human error and human-computer interaction issues''. Since \ac{stpa} was introduced, it is applied to a range of different systems, also in an automotive context. \citet{Sulaman2014a} investigates a forward collision avoidance system using \ac{stpa} and experiences advantages with respect to time effort and covering the dynamic system behavior within the analysis. \cite{Raste2015} uses \ac{stpa} to analyze a fallback strategy for \ac{ad}. \citet{Oscarsson2016} states that other methods are not designed to consider multiple vehicles in the analysis. Therefore, \ac{stpa} is used to analyse a cooperative driving system in \citep{Oscarsson2016}. \citet{Stolte2016} uses the \ac{stpa} to analyse the actuation system of an automated vehicle and proposes a way to better include quasi-continuous control actions in the analysis. \citet{Bagschik2017a} performes an \ac{stpa} on an unmanned protective vehicle for highway road work. 
In \citep{Abdulkhaleq2018a}, \ac{stpa} is applied to identify safety in use requirements for an \ac{ads}. \citet{Abdulkhaleq2018a} finds that the interaction of \ac{ads} with the human, environment or other traffic participants could be sufficiently addressed. \citet{Suo2017} and \citet{Mallya2016} propose ways to integrate \ac{stpa} into the ISO26262 process.

\subsection{Aim of Present Work}
As shown in \autoref{sec:relWorkTof} multiple solutions were developed to address certain problems of teleoperated driving. However, to the knowledge of the authors, there is no literature which attempts to systematically identify the risks and hazards of teleoperated road vehicles. Since this is an important aspect in developing a safety concept, a systematic analysis is performed in the presented work. Due to the advantages in STPA's handling of human flaws and errors, as well as its successful application in different automotive situations, \ac{stpa} is applied to the system at hand.

\section{\uppercase{An Overview of} \acs{stpa}}
\label{sec:stpa}
\begin{figure}[!h]
\newcommand{\myarrow}[1][]{%
	\begin{tikzpicture}[#1]%
	\draw (0,1.85ex) -- (0,0) -- (0.45em,0.0em);
	\draw (0.25em,0.2em) -- (0.45em,0.0em) -- (0.25em,-0.2em);
	\end{tikzpicture}%
}
\centering
\begin{tikzpicture}
\def\desWidth{6.8cm}%
\def\innerSep{0.1cm}
\def\numberOfNodes{4}
\def\rightOf{0.08cm}

\tikzset{
	arrow/.style={
		draw,
		minimum height=1.7cm,
		inner sep=\innerSep,
		align = center,
		text width = 1/\numberOfNodes*\linewidth - (\numberOfNodes+1)*\innerSep -0.08cm,
		shape=signal,
		signal from=west,
		signal to=east,
		signal pointer angle=150,
	},
	bracket/.default=0.2cm,
	bracket/.style={to path={ ([yshift=0.05cm]\tikztostart) -- ([yshift=#1]\tikztostart) -- ([yshift=#1]\tikztotarget) \tikztonodes -- ([yshift=0.05cm]\tikztotarget)}},
	outcome/.style={rectangle, inner sep = 0cm, minimum height=0.5cm, align = left}
}

\node[arrow](step1) {\scriptsize {\bf Step 1:} Purpose of Analysis};
\node[arrow,right = \rightOf of step1](step2) {\scriptsize {\bf Step 2:} Control Structure};
\node[arrow, right = \rightOf of step2](step3) {\scriptsize {\bf Step 3:} Unsafe Control A.};
\node[arrow, right = \rightOf of step3](step4) {\scriptsize {\bf Step 4:} Causal Factors};

\draw[] (step1.north west) to[bracket] node[above=.01em] {\scriptsize Preparation} (step2.north east);
\draw[] (step3.north west) to[bracket] node[above=.01em] {\scriptsize Main Analysis} (step4.north east);

\def\OutSize{\tiny}
\node[outcome, below = 0.05cm of step1.south west, anchor = north west](outStep1){\OutSize \myarrow \ Losses \\ \OutSize \myarrow \ System Hazards \\ \OutSize \myarrow \ System Constr.};

\node[outcome, below = 0.05cm of step2.south west, anchor = north west](outStep2){\OutSize \myarrow \ Control Structure \\ \OutSize  \\ \OutSize };

\node[outcome, below = 0.05cm of step3.south west, anchor = north west](outStep3){\OutSize \myarrow \ \acs{uca} \\ \OutSize \myarrow \ Controller Constr. \\ \OutSize };

\node[outcome, below = 0.05cm of step4.south west, anchor = north west](outStep4){\OutSize \myarrow \ Loss Scenarios \\ \OutSize  \\ \OutSize };

\end{tikzpicture}
\vspace{0.1em}
\caption{Overview on the \acs{stpa} process and the outcome of the individual steps.} \label{fig:stpaDescr}
\end{figure}
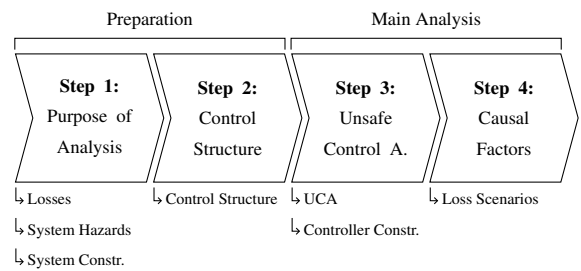

\noindent Before applying the \ac{stpa} to the teleoperation system in \autoref{sec:stpaAppl}, an overview of this method is given. The \ac{stpa} is divided into four steps, as shown in \autoref{fig:stpaDescr}. While step 1 and step 2 are considered as preparation, steps 3 and 4 make up the main analysis of the system. The most important aspects of the individual steps are provided in \autoref{sec:DesStep1} to \autoref{sec:DesStep4}. For further information on \ac{stpa}, refer to \citep{Leveson2018}.

\subsection{Defining the Purpose of the Analysis}
\label{sec:DesStep1}
The first step of performing an \ac{stpa} involves defining the purpose of the analysis, which involves losses, system boundaries, system-level hazards and \acp{sc}. According to \citet[p. 16]{Leveson2018}, a loss could be anything of value to a stakeholder, while a hazard is: ``A system state or set of conditions that, together with a particular set of worst-case environmental conditions, will lead to an accident (loss)'' \citep[p. 184]{Leveson2011}. To define the system-level hazards, the system and its boundaries need to be determined. Finally, the \acp{sc} are identified. According to \citet{Leveson2018}, they can be simply formulated by inverting the hazards or by specifying what needs to happen if a hazard occurs.
\subsection{Control Structure}
Modeling the system as a control structure is a central aspect of \ac{stpa}, that formulates safety as a control problem. Therefore, the next step of the \ac{stpa} requires generating a hierarchical control structure of the system. At a minimum, the control structure consists of a controller and a controlled process. The controller has some control authority over the controlled process by \acp{ca} and receives feedback from the controlled process. The controller has some internal control algorithm which calculates and provides the \ac{ca}. The process model of the controller represents the internal beliefs of the controller, for example, about the controlled process or the environment. \citep[p. 22-25]{Leveson2018}\\

\subsection{Identifying Unsafe Control Actions}
\label{sec:stpaStep3}
After the control structure is developed, the main analysis starts with identifying \acp{uca}. ``An Unsafe Control Action (UCA) is a control action that, in a particular context and worst-case environment, will lead to a hazard'' \citep[p. 35]{Leveson2018}. To identify \acp{uca}, every \ac{ca} is analyzed with respect to whether providing, not providing, providing too long/short or providing too early/late could lead to one of the hazards identified in step 1. It is important to specify a context which makes the \ac{ca} unsafe. According to \citet[p. 36]{Leveson2018}, a context could be an environmental condition, state of the controlled process, or previous actions or parameters. \citet{Thomas2013} extends the \ac{stpa} through systematic means to identify context variables. Accordingly, he identified process model variables that represent the information or beliefs a controller requires about the controlled process or the environment to provide a \ac{ca}. \citet{Thomas2013} derived the system level variables from the system level hazards. As a next step, discrete values are assigned to the variables. To identify \acp{uca}, different combinations of the values\textemdash the context\textemdash are checked, if providing or not providing the \ac{ca} in this context can lead to a hazard. Controller constraints can further be derived based on the \ac{uca}.

\subsection{Identifying Causal Factors}
\label{sec:DesStep4}
The final step of \ac{stpa} is to identify the potential causes of unsafe behavior. According to \citet[p. 169]{Thomas2013}, safety constraints can be violated either by a controller providing a \ac{uca} (case 1) or by an appropriate \ac{ca} not being followed (case 2). To identify potential causes for the first case, the entire feedback path, including the controller itself (process model, control algorithm) needs to be analyzed. Potential external influences or communications to other controllers also need to be considered. To find causes for appropriate \ac{ca} not being followed, the \ac{ca} path needs to be analyzed, including actuators, the controlled process, disturbances, environmental influences or other controllers. \citet[p. 169]{Thomas2013} provides a classification of \acp{cf}, which can be used as guidance to analyze the control structure.

\section{\uppercase{System Description}}
\label{sec:systemDescription}
\begin{figure}[!h]
\centering
\begin{tikzpicture}[auto, node distance=1.5cm,>=latex']

\def\desWidth{6.8cm}%
\def\horDist{0.3cm}%
\def\verDist{0.1cm}
\def\verDistSmall{1.2cm}
\def\minimumHeight{0.8cm}

\def\desHeight{2.5cm}

\tikzset{
	component/.style={draw,rectangle,rounded corners, minimum height=0.76cm},
	vehicle/.style={rectangle,rounded corners, minimum height=\desHeight},
	interface/.style={draw,rectangle,rounded corners, minimum height=\desHeight},
	container/.style={draw, rectangle,dashed,inner sep=0.15cm,minimum height=1cm},
	container2/.style={rectangle,dashed,inner sep=0.15cm,minimum height=1cm},
	controlAction/.style={draw, thick,->,>=stealth},
	feedback/.style={draw, thick,->,>=stealth},
	line/.style={draw, thick,-,>=stealth},
	Doublearrow/.style={draw, thick,<->,>=stealth},
}

	\def\nodeWidth{0.5cm}

	\node [vehicle, minimum width = \nodeWidth*3] (vehicle) {\includegraphics[width=\nodeWidth*3]{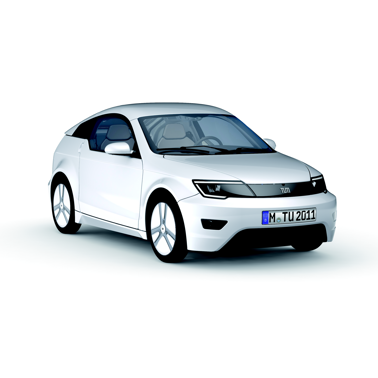}};
		
	\node [component, right = \horDist-0.4cm of vehicle.north east, anchor = north west, minimum width = \nodeWidth] (sensor) {\includegraphics[width=\nodeWidth]{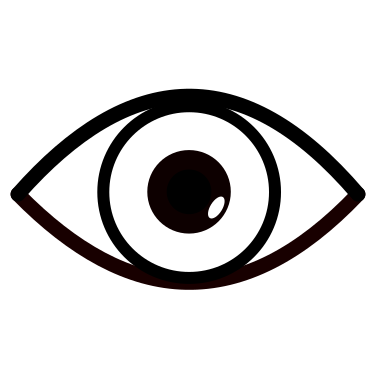}};
	
	\node [component, below = \verDist of sensor, minimum width = \nodeWidth] (engine) {\includegraphics[width=\nodeWidth]{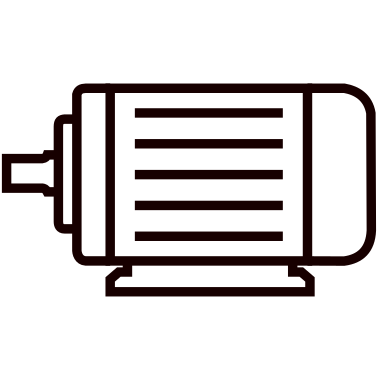}};
	
	\node [component, below = \verDist of engine, minimum width = \nodeWidth] (swVeh) {\includegraphics[width=\nodeWidth]{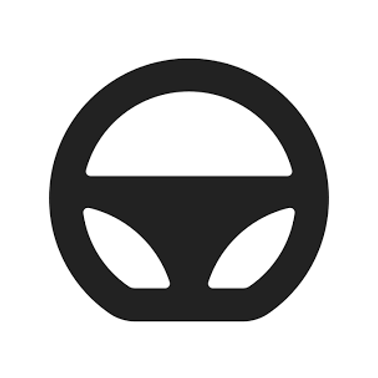}};

	\begin{scope}[on background layer]
		\node [container,fit=(vehicle)(sensor)(swVeh)(engine)] (vehicleComp) {};
		\node at (vehicleComp.north) [fill=white] {\text{\scriptsize Vehicle}};
	\end{scope}

	\node [vehicle, right = \horDist of sensor.north east, anchor = north west, minimum width = \nodeWidth*2] (net) {\includegraphics[width=\nodeWidth+0.4cm]{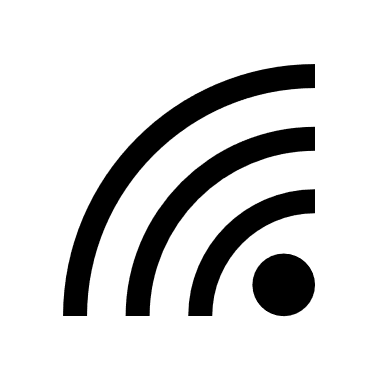}};
	
	\begin{scope}[on background layer]	
	\node [container2,fit=(net)] (operComp) {};	
	\node at (operComp.north) [fill=white] {\text{\scriptsize 4G/5G}};
	\end{scope}

	\node [interface, right = \horDist of net.north east, anchor = north west, minimum width = \nodeWidth] (interface) {};
	
	\node[align=center,rotate=90] at (interface.center) {\small Interface};
	
	\node [component, right = \horDist of interface.north east, anchor = north west, minimum width = \nodeWidth] (display) {\includegraphics[width=\nodeWidth]{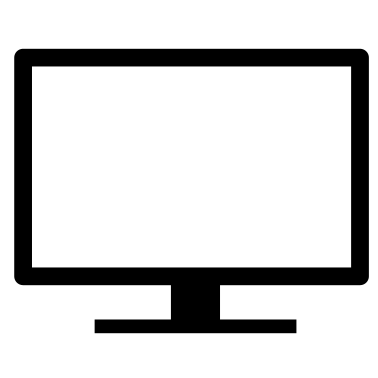}};
	
	\node [component, below = \verDist of display, minimum width = \nodeWidth] (pedal) {\includegraphics[width=\nodeWidth]{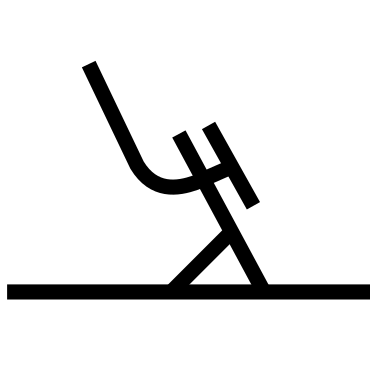}};
	
	\node [component, below = \verDist of pedal, minimum width = \nodeWidth] (swOp) {\includegraphics[width=\nodeWidth]{Pictures/sw_.png}};
	
	\node [bob, mirrored, right = \horDist of pedal, minimum width = \nodeWidth+0.2cm] (operator) {\scriptsize Operator};
	
	\begin{scope}[on background layer]	
		\node [container,fit=(interface)(display)(pedal)(swOp)(operator)] (operComp) {};	
		\node at (operComp.north) [fill=white] {\text{\scriptsize Control Center}};
	\end{scope}
	
\draw [controlAction] (interface.east|-display.west) -- (display.west); 
\draw [controlAction] (pedal.west) -- (interface.east|-pedal.west); 
\draw [controlAction] (swOp.west) -- (interface.east|-swOp.west); 
\draw [controlAction] (operComp.west|-swOp.west) -- (vehicleComp.east|-swOp.west) node[midway, below] {\tiny control signals}; ; 

\draw [controlAction] (vehicleComp.east|-display.west) -- (operComp.west|-display.west) node[midway, below] {\tiny feedback}; 

\end{tikzpicture}

\caption{Overview of the teleoperation system} \label{fig:sysDes}
\end{figure}

\noindent In this section, the system to be analyzed within the present work is discussed. An overview is shown in \autoref{fig:sysDes}. The vehicle perceives its environment using camera sensors. This sensor information, together with the vehicle's internal states, is sent to the control center. A virtual representation of the vehicle's surroundings is generated within the interface \citep{Georg2019a}. This information is provided to the operator using displays. Based on the feedback, the operator can provide control signals using a steering wheel, throttle and brake pedal actuators. A vehicle steering wheel angle and desired velocity are calculated based on these inputs. The feedback, as well as the control signals, are transmitted to the vehicle using the cellular network. A detailed overview of the individual components, including latencies within the actuators and sensors chain, is published by \citet{Georg2020a}.

\vspace{-0.5cm}

\section{\uppercase{Application of STPA on Teleoperation System}}
\label{sec:stpaAppl}
\noindent In the following sections the individual steps, introduced in \autoref{sec:stpa}, are applied to the teleoperation system in \autoref{sec:systemDescription}.
\subsection{Defining the Purpose of the Analysis}
\label{sec:step1}
We identified the following stakeholders to the teleoperation system: vehicle passengers, other traffic participants and property owners. After identifying the stake or value of each stakeholder, we determined the losses L-1 to L-2. \citet[p. 148]{Leveson2018} provided some losses and hazards for automotive industry. Since these did fit for our system, we borrowed L-1, L-2, H-1 and H-5 from \citet[p. 148]{Leveson2018}.
\begin{enumerate}[labelindent=0pt,align=left, labelwidth=\widthof{\ref{last-itemL}}, label=L-\arabic*,leftmargin=!]
	\item Loss of life or injury to people
	\item Damage to ego vehicle or objects outside the ego vehicle
\end{enumerate}

\noindent If the system in \autoref{sec:systemDescription} is part of a taxi fleet, the service provider could be a stakeholder with its own goals. The operator or an \ac{oem} may also have some stake in the system. An \ac{oem} could be concerned about its image, and ``Loss of \ac{oem} image'' could be another loss. The stake of vehicle passengers could also be comfort or punctuality. However, only L-1 to L-2 are considered within the scope of this work.

\noindent The system-level hazards are identified in the next step. Following the ideas of \citet{Leveson2018}, \ref{first-itemH} to \ref{last-itemH} could lead to a loss under some worst-case environmental condition:

\begin{enumerate}[labelindent=0pt, align=left,labelwidth=\widthof{\ref{last-itemH}}, label=H-\arabic*,leftmargin=!]
	\item System does not maintain safe distance from nearby objects [L\=/1, L\=/2] \label{first-itemH}
	\item System leaves intended lane [L\=/1, L\=/2]
	\item System behavior is breaking the law (e.g., red lights, stop sign) [L\=/1, L\=/2]
	\item Vehicle exceeds safe operating envelope for environment (speed, lateral/longitudinal forces) [L\=/1, L\=/2] \label{last-itemH}
\end{enumerate}

\noindent The losses that individual hazards could cause are specified in brackets. The formulated system-level hazards and losses are very general and abstract. However, this is intended by \citet{Leveson2018}. The hazards should not be on a component level and no causes for hazards should be part of the hazard description. The causes for hazards on a component level are investigated in the following \ac{stpa} steps. Unnecessary detail should be omitted to better identify missing aspects and keep the list manageable \citep[p. 19]{Leveson2018}. \citet{Leveson2018} propose refining the hazards into sub-hazards in a later step if required. 
\acp{sc} can be identified based on the identified hazards. Only two examples are provided within the scope of this work.

\begin{enumerate}[labelindent=0pt,align=left, labelwidth=\widthof{\ref{last-itemSC}}, label=SC-\arabic*,leftmargin=!]
	\item The system must always be able to react on obstacles [H\=/1]
	\item If the system exceeds its dynamic boundaries, this needs to be detected and countermeasures need to be taken [H\=/4] \label{last-itemSC}
\end{enumerate}

\subsection{Control Structure}
\label{sec:step2}
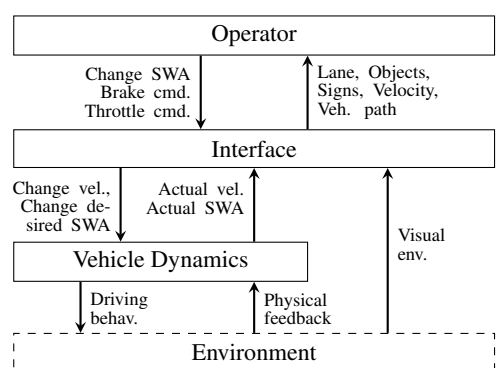
\begin{figure}[!h]
\centering
\begin{tikzpicture}[auto, node distance=1.2cm,>=latex']
\tikzset{
	block/.style = {draw, fill=white, rectangle, minimum height=0.50cm, minimum width=18em},
	blockDashed/.style = {draw, dashed, fill=white, rectangle, minimum height=0.5cm, minimum width=18em},
	blockDotted/.style = {draw, dotted, fill=white, rectangle, minimum height=0.5cm, minimum width=18em},
	arrow/.style={draw, thick,->,>=stealth},
	line/.style={draw, thick,-,>=stealth},
	Doublearrow/.style={draw, thick,<->,>=stealth},
	container/.style={draw, rectangle,dotted,inner sep=0.15cm,minimum height=1cm},
}

	
	\node [block] (operator) {\small Operator};
	\node [block, below = of operator.west, yshift =-0.3cm, anchor = west] (hmi) {\small Interface};

	\node [block, minimum width = 11em, yshift =-0.3cm, below = of hmi.west, anchor = west] (veh) {\small Vehicle Dynamics};
	

	\node [blockDashed, minimum width = 18em, below = of veh.west, anchor = west] (env) {\small Environment};

	\draw [arrow] ([xshift=-2em] operator.south) -- ([xshift=-2em] hmi.north) node[midway,left, text width = 6em, align = right, yshift = -0.08cm] {\begin{spacing}{0.7}
		\scriptsize Change \acs{swa}\\ Brake cmd.\\ Throttle cmd.
		\end{spacing} };
	\draw [arrow] ([xshift=2em] hmi.north) -- ([xshift=2em] operator.south) node[midway,right,text width = 5em, yshift = -0.08cm] {\begin{spacing}{0.7}
		\scriptsize Lane, Objects,\\ Signs, Velocity,\\ Veh. path
		\end{spacing}  };
	\draw [arrow] ([xshift=-5em]hmi.south) -- ([xshift=-5em]hmi.south|-veh.north) node[midway,left,text width = 4.5em,yshift=-0.1cm, align=right] {\begin{spacing}{0.7}
		\scriptsize Change vel.,\\ Change desired \acs{swa}
		\end{spacing}};
	

	\draw [arrow] ([xshift=3.5em]veh.north) -- ([xshift=3.5em]veh.north|-hmi.south) node[midway,left,align = right ,text width = 8em] {\begin{spacing}{0.7}
		\scriptsize Actual vel.\\ Actual \acs{swa}
		\end{spacing}};
	\draw [arrow] ([xshift=5em]env.north) -- ([xshift=5em]env.north|-hmi.south) node[midway,right,text width = 4em] {\begin{spacing}{0.7}
		\scriptsize Visual\\ env.
		\end{spacing}};
	\draw [arrow] ([xshift=-3.0em]veh.south) -- ([xshift=-3.0em]veh.south|-env.north) node[midway, right,yshift=-0.1cm,text width = 4em] {\begin{spacing}{0.7}
		\scriptsize Driving behav.
		\end{spacing}};
	\draw [arrow] (env.north) -- (env.north|-veh.south) node[midway,right,yshift=-0.1cm,text width = 4em] {\begin{spacing}{0.7}
		\scriptsize Physical feedback
		\end{spacing}};

	

\end{tikzpicture}

\caption{\ac{stpa} control structure of the teleop. system} \label{fig:mainCS}
\end{figure}

An \ac{stpa} control structure for the teleoperation system presented in \autoref{sec:systemDescription} is shown in \autoref{fig:mainCS}.

\noindent The Operator provides \acp{ca}, such as the change \ac{swa} command, brake command and throttle command to the interface. Thus, only the primary driving tasks \citep{Bubb2003}
are considered for the analysis. According to \citet{Stolte2016} ``change \ac{swa}'' instead of the continous \ac{swa} is used. However, ``hold \ac{swa}'' was not included as a separate command, but considerd as ``not providing'' the ``change \ac{swa}'' command in \autoref{sec:step3}. The operator receives visual feedback from the interface. The display and input devices are not part of the control structure to reduce complexity. They are, together with sensors and actuators, added in step 4 of the analysis. The human operator is modeled as a controller in the provided control structure.
Therefore, the human operator can be included in the analysis and human flaws or errors can be identified which is one of the advantages of using \ac{stpa}. The human operator also has some control algorithm and process models. \citet{Rasmussen1983} proposed the three levels of performance of skilled human operators. This is a model describing the cognitive processes of humans from input to actions. The model distinguishes between three different layers: skill-based behavior, rule-based behavior and knowledge-based behavior. In \citep{Donges2009}, these layers are assigned to the different layers of driving tasks (stabilization, guidance and navigation) proposed by \citet{Donges1982}. Therefore, we consider this model a control algorithm for the human operator in context of a driving task. The process model of a human operator is called a mental model \cite{Thomas2015a}.

\citet{Rasmussen1983} states that the different mental models humans create are the reason for human performance in coping with complexity. Since the operator is not located inside the vehicle, the mental model is largely dependent and updated by the received feedback and the presentation of the feedback. \\
Based on the operator \acp{ca}, the interface calculates a change velocity command and a change desired \ac{swa} command. The interface does also contain certain beliefs about the controlled process, such as the steering ratio, sensor positions, etc. These beliefs make up the process model for the interface. The interface receives feedback from the physical vehicle and the vehicle's environment, creates a scene representation of the environment and provides this information to the operator. Further information on the interface is published by \citet{Georg2019a}. The environment is not part of the system itself. Nevertheless, we decided to include it in the control structure, similar to \citet{Placke2015}, to indicate where certain information comes from.\\
The desired velocity and \ac{swa} is transmitted to the vehicle via wireless network. Even if the actual signal sent to the vehicle is a desired velocity, it can be advantageous for the analysis to split the \ac{ca} into increase and decrease velocity. The network is not visualized within \autoref{fig:mainCS}. Similar to sensors and actuators, the information about the network is neglected in this step. What happens to the \ac{ca} on its way to the controlled process is part of step 4 during the analysis.\\
The presented control structure is an abstract representation of the real system. The control structure in \autoref{fig:mainCS} is not dependent on any implementation details or component decisions and is therefore valid for a variety of different teleoperated road vehicles.
The vehicle in \autoref{fig:mainCS} is only represented by its dynamic behaviour. The reason is to perform some abstraction and simplification. From this, we cannot perform a detailed analysis on the vehicle internal control loops. However, we can still consider the interaction of the vehicle and other entities since the input and outputs stay the same. The vehicle control loop is not intrinsic to the teleoperation concept itself. \citet{Stolte2016} performed an \ac{stpa} solely on the vehicle actuation system in the context of \ac{ad}.
The goal of this analysis is to investigate conceptual problems of teleoperated driving in a first step and not some implementation or hardware specific hazards. We are also not including certain available safety measures (\autoref{sec:relWorkTof}), to not overlook important aspects or better solutions during the analysis. The fact that \ac{stpa} is a top-down approach allows to make an analysis before specific components and design decisions are made. The results can therefore be considered in the later development.

\subsection{Identifying Unsafe Control Actions}
\label{sec:step3}

Applying the approach of \citet{Thomas2013}, we ended up with the process model variables in \autoref{tab:contexts}, making up the contexts. In reality, road networks or traffic regulations are more complex, especially when considering urban situations. However, to reduce complexity for a first analysis, this abstract representation was chosen.
\begin{table}[h]
\caption{Process model variables making up the contexts}\label{tab:contexts} \centering 
\begin{tabularx}{0.45\textwidth}{|X|X|} 
\hline
Conditions & Values \\
\hline 
Vehicle motion& 
	- Stopped \newline
	- Moving \\

\hline 
Traffic participants relative to ego vehicle&
	- None \newline
	- Same lane in front \newline
	- Same lane behind \newline
	- Neighboring lane \\

\hline
Road Surface&
	- High $\mu$ \newline
	- Low $\mu$ \\

\hline
Regulatory elements (signs, lights, etc.)&
	- Yes \newline
	- No \\

\hline
Lane&
	- $\dot{\kappa} \neq 0$  \newline
	- $\dot{\kappa} = 0$ \\

\hline 
\end{tabularx} 
\end{table}
\\Due to the high amount of identified \acp{uca}, only the \acp{uca} related to the brake command are presented in \autoref{tab:ucaOp}. If a certain condition is not explicitly mentioned within the \ac{uca} description, all its values could lead to the specified hazard. The controller constraints, resulting from \acp{uca}, are not explicitly mentioned here.

\begin{table*}[htb]
	\caption{Unsafe Control Actions for Brake Command.} \label{tab:ucaOp}\centering	
	\begin{tabularx}{\textwidth} {|X|p{4.8cm}|X|X|} 
	\hline 
	Not providing causes hazards
	& Providing causes hazards 
	& To early, too late, out of order 
	& Stopped too soon, applied too long \\
		
	\hline 

	UCA-1: O. does not provide \acs{sBp}, if vehicle is moving and object is in/approaching same lane to the front [H-1] \newline
	UCA-2: O. does not provide \acs{sBp}, if vehicle is moving and regulatory elements are present [H-3] \newline
	UCA-3: O. does not provide \acs{sBp}, if vehicle is moving on low $\mu$ and $\dot{\kappa} \neq 0$ [H-4, H-2]\newline
	UCA-4: O. does not provide \acs{sBp}, if vehicle is moving on low $\mu$, $\dot{\kappa} \neq 0$ and objects in/approaching neighboring lane [H-1] 
	&UCA-5: O. provides excessive \acs{sBp}, if vehicle is moving and object in/approaching same lane to the rear [H-1] \newline
	UCA-6: O. provides excessive \acs{sBp}, if vehicle is moving and no obstacle in/approaching same lane to the front [H-3] \newline
	UCA-7: O. provides insufficient or excessive \acs{sBp}, if vehicle is moving on low $\mu$ and $\dot{\kappa} \neq 0$ [H-4, H-2] \newline
	UCA-8: O. provides excessive \acs{sBp}, if vehicle is moving on low $\mu$ and $\dot{\kappa} \neq 0$ and an obstacle is in/approaching neighboring lane [H-1, H-2] \newline
	UCA-9: O. provides insufficient \acs{sBp}, if vehicle is moving and object is in/approaching same lane to the front [H-1] \newline
	UCA-10: O. provides insufficient \acs{sBp}, if vehicle is moving and regulatory elements are present [H-3] \newline

	& UCA-11: O. provides \acs{sBp} too early, if vehicle is moving and object in/approaching same lane to the rear [H-1]\newline
	UCA-12: O. provides \acs{sBp} too late, if vehicle is moving and object in/approaching same lane to the front [H-1]\newline
	UCA-13: O. provides \acs{sBp} too late, if vehicle is moving and regulatory elements are present [H-3] \newline
	UCA-14: O. provides \acs{sBp} too late, if vehicle is moving on low $\mu$ and $\dot{\kappa} \neq 0$ [H-2, H-4] \newline
	& UCA-15: O. stops providing \acs{sBp} to soon, if vehicle is moving and object in/approaching lane to the front [H-1]\newline
	UCA-16: O. stops providing \acs{sBp} to soon, if vehicle is moving and regulatory elements are present [H-3]\newline
	UCA-17: O. stops providing \acs{sBp} to soon, if vehicle is moving on low $\mu$ and $\dot{\kappa} \neq 0$ [H-2, H-4] \\ 
		
	\hline 
\end{tabularx}
\end{table*} 

\subsection{Identifying Causal Factors}
\label{sec:step4}
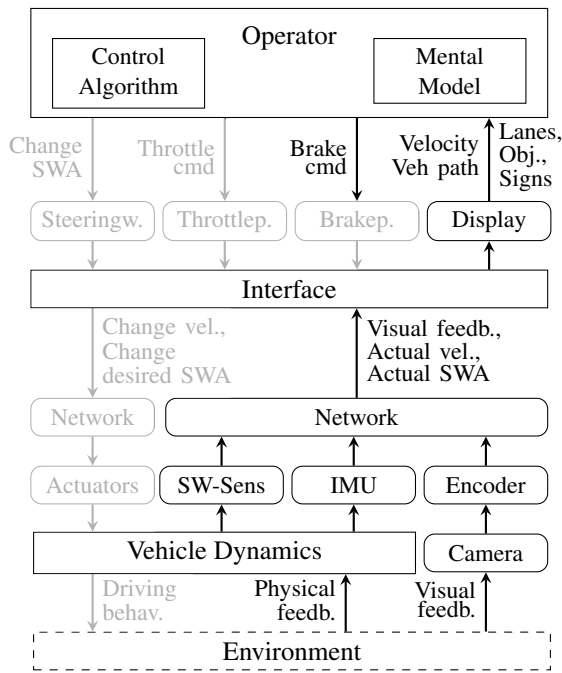
\begin{figure}[!h]
\centering
\begin{tikzpicture}[auto, node distance=1.5cm,>=latex']

\def\desWidth{6.8cm}%
\def\horDist{0.1cm}%
\def\verDist{1.6cm}
\def\verDistSmall{0.9cm}
\def\minimumHeight{0.5cm}

\tikzset{
	controller/.style = {draw, fill=white, rectangle, minimum height=\minimumHeight, minimum width=\desWidth},
	process/.style = {draw, fill=white, rectangle, minimum height=\minimumHeight, minimum width=\desWidth},
	environment/.style = {draw, dashed, fill=white, rectangle, minimum height=\minimumHeight, minimum width=\desWidth},
	sensor/.style={draw,rectangle,rounded corners, minimum height=\minimumHeight},
	actuator/.style={draw,rectangle, rounded corners, minimum height=\minimumHeight},
	container/.style={draw, rectangle,dotted,inner sep=0.23cm,minimum height=\minimumHeight},
	controlAction/.style={draw, thick,->,>=stealth},
	feedback/.style={draw, thick,->,>=stealth},
	line/.style={draw, thick,-,>=stealth},
	Doublearrow/.style={draw, thick,<->,>=stealth},
}

	\def\nodeWidth{\desWidth*0.25 - \horDist*0.75}

	\node [controller] (operator) {
	\begin{tikzpicture}[auto, node distance=1.0cm,>=latex']	
		\node [rectangle, minimum height=0.1cm, minimum width=2cm] (op){Operator};
		\node [draw,rectangle, minimum height=0.2cm, minimum width=2cm, right = 0.1cm of op.east, anchor=north west, text width = 4em, align=center] (processModel){\footnotesize Mental Model};
		\node [draw,rectangle, minimum height=0.2cm, minimum width=2cm, left = 0.1cm of op.west, anchor=north east, text width = 4em, align=center] (controlAlgorithm){\footnotesize Control Algorithm};

	\end{tikzpicture}
	};	
	\node [actuator, below = \verDist-0.5cm of operator.south west, anchor = north west, minimum width = \nodeWidth] (swOp) {\footnotesize Steeringw.};	
	\node [actuator, right = \horDist of swOp.east, anchor = west, minimum width = \nodeWidth] (tPedOp) {\footnotesize Throttlep.};	
	\node [actuator, right =  \horDist of tPedOp.east, anchor = west, minimum width = \nodeWidth] (bPedOp) {\footnotesize Brakep.};	
	\node [sensor, right =  \horDist of bPedOp.east, anchor = west, minimum width = \nodeWidth] (dispOp) {\footnotesize Display};	
	\node [controller, below = \verDistSmall of swOp.west, anchor = west] (hmi) {Interface};	
	\node [actuator, below =\verDist+0.1cm of hmi.west, anchor = west, minimum width = \nodeWidth] (netDL) {\footnotesize Network};	
	\node [actuator, below =\verDistSmall of netDL.west, anchor = west, minimum width = \nodeWidth] (Actuator) {\footnotesize Actuators};
	\node [sensor, below =\verDist+0.1cm of hmi.east, anchor = east, minimum width = 3.0*\nodeWidth] (netUL) {\footnotesize Network};	
	\node [sensor, below =\verDistSmall of netUL.east, anchor = east, minimum width = \nodeWidth] (encoder) {\footnotesize Encoder};
	\node [sensor, right =-\horDist of encoder.west, anchor = east, minimum width = \nodeWidth] (imu) {\footnotesize IMU};
	\node [sensor, right =-\horDist of imu.west, anchor = east, minimum width = \nodeWidth] (dSens) {\footnotesize SW-Sens};	
	\node [sensor, below =\verDistSmall of encoder.east, anchor = east, minimum width = \nodeWidth] (cam) {\footnotesize Camera};
	\node [process, right =-\horDist of cam.west, anchor = east, minimum width =3*\nodeWidth] (veh) {Vehicle Dynamics};
	\node [environment, below = \verDist- 0.3cm of veh.west, anchor = west] (env) {Environment};	
	
	\draw [controlAction] (operator.south-|swOp.north) -- ( swOp.north) 
	node[midway,left,align = right,text width =3.0em] {\vspace{0.15cm}\begin{spacing}{0.7}\footnotesize Change\\ \ac{swa}	\end{spacing} };
	\draw [controlAction] (swOp.south) -- ( swOp.south|-hmi.north);
	\draw [controlAction] (operator.south-|bPedOp.north) -- (bPedOp.north) 
	node[midway,left,align = right,text width =3.0em] {\vspace{0.15cm}\begin{spacing}{0.7}\footnotesize Brake\\ cmd	\end{spacing}};
	\draw [controlAction] (bPedOp.south) -- ( bPedOp.south|-hmi.north);
	\draw [controlAction] (operator.south-|tPedOp.north) -- (tPedOp.north) 
	node[midway,left,align = right,text width =3.0em] {\vspace{0.15cm}\begin{spacing}{0.7}\footnotesize Throttle\\ cmd	\end{spacing}};
	\draw [controlAction] (tPedOp.south) -- ( tPedOp.south|-hmi.north);		
	\draw [controlAction] (hmi.south-|netDL.north) -- ( netDL.north) 
	node[midway,right,text width =5em] {\vspace{0.15cm}\begin{spacing}{0.7}\footnotesize Change vel., \\ Change \\ desired \ac{swa}\end{spacing}};
	\draw [controlAction] (netDL.south) -- ( netDL.south|-Actuator.north);	
	\draw [controlAction] (Actuator.south) -- ( Actuator.south|-veh.north);
	\draw [controlAction] (netDL.south|-veh.south) -- (netDL.south|-env.north) node[midway,right, text width =3.0em] {\vspace{0.15cm}\begin{spacing}{0.7}\footnotesize Driving behav.	\end{spacing}};		
	\draw [feedback] (dispOp.north) -- (dispOp.north|-operator.south)
	node[midway,right, text width =3em] {\vspace{0.15cm}\begin{spacing}{0.7}\footnotesize Lanes, Obj., Signs \end{spacing}} node [midway, left, align=right ,text width =3.6em] {\vspace{0.15cm}\begin{spacing}{0.7}\footnotesize Velocity\\ Veh path \end{spacing}};
	\draw [feedback] (hmi.north-|dispOp.south) -- (dispOp.south);	
	\draw [feedback] (cam.north) -- (encoder.south);
	\draw [feedback] (env.north-|cam.south) -- (cam.south) node[midway,left, text width =2.6em, align = right] {\vspace{0.15cm}\begin{spacing}{0.7}\footnotesize Visual feedb.\end{spacing}};	
	\draw [feedback] (imu.north) -- (imu.north|-netUL.south);
	\draw [feedback] (dSens.north) -- (dSens.north|-netUL.south);
	\draw [feedback] (encoder.north) -- (encoder.north|-netUL.south);
	\draw [feedback] (veh.north-|imu.south) -- (imu.south);
	\draw [feedback] (veh.north-|dSens.south) -- (dSens.south);	
	\draw [feedback] (netUL.north) -- (netUL.north|-hmi.south) node[midway,right, text width =5.5em] {\vspace{0.15cm}\begin{spacing}{0.7} \footnotesize Visual feedb.,\\ Actual vel.,\\ Actual \ac{swa}	\end{spacing}};	
	\draw [feedback] ([xshift=-0.5*\nodeWidth]env.north-| veh.south east) -- ([xshift=-0.5*\nodeWidth]veh.south east) node[midway,left, text width =4.0em, align=right] {\vspace{0.15cm}\begin{spacing}{0.7} \footnotesize Physical feedb.	\end{spacing}};
	
%
	\fill[fill=white, opacity=.70] (operator.south west) -- (operator.south-|tPedOp.north) -- (bPedOp.north west)  -- (bPedOp.north east)  -- (bPedOp.south east)-- (bPedOp.south|-hmi.north) -- (hmi.north west)-- (hmi.south west)--([xshift=1.0cm]hmi.south-|netDL.east)--([xshift=1.0cm] netDL.north east)-- (netDL.north east) --(veh.north-|netDL.east) -- (veh.north west)-- (veh.south west)-- ([xshift=0.5cm]veh.south-|netDL.east) --([xshift=1.0cm]env.north-|netDL.east)-- ([xshift = -0.3cm]env.north west) -- ([xshift = -0.3cm]operator.south west) -- cycle;

\end{tikzpicture}

\caption{Control Structure of the teleoperation system} \label{fig:detCS}
\end{figure}

To identify potential causes for the \acp{uca} provided in \autoref{tab:ucaOp}, how the feedback is provided to the controllers and how control actions are executed need to be considered. The control structure is updated accordingly in \autoref{fig:detCS} by including sensors, actuators, network, mental model and control algorithm. Entities that are not explicitly addressed within this chapter are grayed out.

\noindent To identify \acp{cf} for UCA-1 an examination of causes in the highlighted entities within \autoref{fig:detCS} is required. Starting with the operator who initially provided UCA-1, the operator's control algorithm and mental model need to be considered potential causes. The operator might have multiple mental models to represent the environment, the interface and the vehicle. One reason for UCA\=/1 could be an inconsistent, incomplete or incorrect mental model, which does not (completely) represent the reality. This is often referred to as situation awareness \citep[p. 188]{Leveson2018}. When thinking about the information the operator requires to avoid UCA-1, the following \acp{cf} regarding the mental model can be formulated:

\begin{enumerate}[labelindent=0pt, align=left,labelwidth=\widthof{\ref{last-item}}, label=CF-\arabic*,leftmargin=!]
	\item Mental model contains no/wrong information about surrounding objects and their relative position to the ego vehicle.\label{cf:firstFBpath}
	\item Mental model contains no/wrong information about the motion or dimensions of the ego vehicle.
	\item Mental model contains no/wrong information about vehicle/actuator/interface behaviour.
\end{enumerate}

\noindent In a next step, we can identify further reasons for the above mentioned \acp{cf}. The mental models are constantly updated by inputs, training or experience \citep[p. 185]{Leveson2018}. Potential reasons for the above mentioned \acp{cf} could be:

\begin{enumerate}[resume,labelindent=0pt,align=left, labelwidth=\widthof{\ref{last-item}}, label=CF-\arabic*,leftmargin=!]
	\item Mental model is not/insufficiently updated due to outer influences of the operator such as distraction.
	\item Mental model is not/insufficiently updated due to insufficient representation of the information (e.g., wrong modality).
	\item Changes in the controlled process (e.g., changing vehicle) results in incorrect mental model.
\end{enumerate}

\noindent Operator visual impairment or health issues could also be a reason. Further \acp{cf} can be identified by analyzing the feedback path in \autoref{fig:detCS} from environment (bottom) to operator (top). Starting with visual feedback the following \acp{cf} are identified:

\begin{enumerate}[resume, labelindent=0pt, align=left, labelwidth=\widthof{\ref{last-item}}, label=CF-\arabic*,leftmargin=!]
	\item Object is not within the field of view of the camera sensor. \label{cf:visFeedb1}
	\item Object is obscured by dirt, water or reflections on the lens.
	\item Object is not visible to the camera sensor due to darkness, fog, rain, etc.\label{cf:visFeedbn}
\end{enumerate}

\noindent The feedback of the environment itself cannot be missing or incorrect, since it represents reality. Sensors like the camera, however are only able to capture parts of the reality, which is represented by \ref{cf:visFeedb1} to \ref{cf:visFeedbn}. In case the light reflected by the object reaches the camera sensor, the following \acp{cf} could still occur in the camera or the following encoding step. 

\begin{enumerate}[resume,labelindent=0pt,align=left ,labelwidth=\widthof{\ref{last-item}}, label=CF-\arabic*,leftmargin=!]
	
	\item No or inadequate operation of the camera, encoder, \ac{imu} or \ac{swa}-sensor (hardware or power failure).
	\item Camera, \ac{imu} or \ac{swa}-sensor is not providing output because of connection errors.
	\item Measurement inaccuracies in camera due to color and spatial discretization of the reality could result in the operator being unaware of the object.
	\item Measurement inaccuracies in \ac{imu} or \ac{swa}-sensor leading to wrong operator beliefs about vehicle movement. This could falsely result in no collision within the operator's mental model.
	\item Inaccurate information about distances through lens distortion could lead to wrong distances between the predicted vehicle path and the object within the operator's mental model.
	\item Information loss by image compression of the encoder could result in a hardly detectable object for the operator.\label{last-item}
\end{enumerate}

\noindent The network is one of the most critical parts within the whole teleoperation system. It contributes to the following \acp{cf} of UCA-1:

\begin{enumerate}[resume,labelindent=0pt, labelwidth=\widthof{\ref{last-item}}, label=CF-\arabic*,leftmargin=!]
	\item No network connection at current vehicle location. The feedback about the object is not provided to the operator.
	\item Network is dropping information (packet loss). The feedback about the object is not provided to the operator.
	\item Network inserts wrong information or changes the order of information. Wrong feedback is provided to the operator.
\end{enumerate}

\noindent The interface pre-processing the incoming sensor information also contributes to the list of \acp{cf}. The individual camera streams need to be visualized and placed relative to each other. Therefore, extrinsic and intrinsic camera parameters are required. The interface also provides feedback about the current and future vehicle movement. This information can be calculated e.g., based on the actual \ac{swa} and a dynamic model of the vehicle. This model is part of the interface process model and contains certain beliefs and simplifications about the real vehicle. The same \ac{swa}, for example, results in different vehicle trajectories depending on the vehicle, road friction coefficient or other parameters. 
\begin{enumerate}[resume,labelindent=0pt, labelwidth=\widthof{\ref{last-item}}, label=CF-\arabic*,leftmargin=!]
	\item Interface not working due to power or hardware failure.
	\item Interface uses wrong beliefs about camera position and calibration leading to wrong visualization of objects relative to the vehicle.
	\item Wrong beliefs about vehicle or environmental parameters (friction coefficient $\mu$) results in an incorrect predicted path. Therefore the object could falsely be located outside the predicted path.
\end{enumerate}

\noindent Finally, the display can also be a cause for the operator not (correctly) updating its mental model and therefore resulting in UCA-1:

\begin{enumerate}[resume,labelindent=0pt, labelwidth=\widthof{\ref{last-item}}, label=CF-\arabic*,leftmargin=!]
	\item Display not working due to power or hardware failure
	\item Display reducing information by dropping frames or low resolution
	\item Reflections or dirt on screen masking information	\label{cf:lastFBpath}
\end{enumerate}

\noindent
\ref{cf:firstFBpath} to \ref{cf:lastFBpath} are the \acp{cf} related to the feedback path and the mental model of the operator. But the control algorithm of the operator can also cause UCA-1. As mentioned earlier, the control algorithm of the operator is described using the target-oriented behavior model of \citet{Rasmussen1983}. The driving task the operator is failing to perform in UCA-1 can be assigned to the guidance and stabilization task proposed by \citet{Donges1982}. According to \citet[p. 21]{Donges2009}, knowledge-based, rule-based, and skill-based behavior are utilized to perform those driving tasks. \citet{Donges2009}, however, states that the role of knowledge-based behavior is minimized by routine, which is important due to the long execution times for knowledge-based behavior.

\begin{enumerate}[resume,labelindent=0pt, labelwidth=\widthof{\ref{last-item}}, label=CF-\arabic*,leftmargin=!]
	\item Little routine or training causes the operator to utilize knowledge-based behavior, which could result in a slow or false reaction in response to an object in front of the vehicle.
	\item Wrong routine or training could result in wrong rules being stored, which are later applied as rule-based behavior if a vehicle occurs.
	\item Missing or wrong long-term training could result in a lack of skill-based behavior which is required for dynamic actions in stabilization and guidance tasks.
\end{enumerate}

\noindent
Only by analyzing UCA-1, are 27 \acp{cf} identified that could lead to UCA-1 and therefore to H-1. This procedure needs to be applied to the remaining \acp{uca}. Some of the above identified \acp{cf} are also causes for other \acp{uca}. To reduce the quantity of generated information, we did not add an additional \ac{cf} in this case, but linked the respective \ac{uca} to the existing \ac{cf}. New \acp{cf} are also identified, for example, by analyzing UCA-12 for \acp{cf} that incorporate some timing information.

\begin{enumerate}[resume,labelindent=0pt, labelwidth=\widthof{\ref{last-item}}, label=CF-\arabic*,leftmargin=!]
	\addtocounter{enumi}{8}	
	\item Delayed feedback by camera/encoding because of processing times and discrete operation, which could result in a delayed object detection and reaction.
	\item Delayed feedback information because network delays, which could result in a delayed object detection and reaction.
	\item Processing delays of operator (reaction time), which could result in a delayed object detection and reaction.
\end{enumerate}

\noindent It is also possible that these \acp{cf} have already been identified when analyzing UCA-1, but they should at least be apparent when performing the analysis on UCA-12.\\
Due to the large number of \acp{uca} and \acp{cf} identified using the \ac{stpa}, only a small subset of the results could be presented. Not explicitly mentioned are the \acp{uca} resulting from the interface as well as all the \acp{cf}, except for those related to UCA-1. Also, the \acp{cf} for not or incorrect execution of \ac{ca} and controller constraints need to be identified. The results show the amount of information that is generated using the \ac{stpa}, even after applying different abstractions and simplifications. However, the important aspects, such as operator handling, \acp{uca} related to the operator, the process model variables and a general control structure are presented.

\section{\uppercase{Discussion and Conclusion}}
\label{sec:discussion}
\noindent The present work shows selected results of applying \ac{stpa} to a teleoperated road vehicle. Previous publications often addressed single aspects that affected the safety of teleoperated road vehicles and proposed solutions. To obtain a deeper understanding of the safety challenges inherent to teleoperated road vehicles, we decided to perform a thorough analysis. However, for a first analysis, some simplifications and abstractions were performed. The operator performed only primary driving tasks and the vehicle feedback only consisted of camera images, velocity and \ac{swa}. The vehicle was also not divided into single components and control loops. This however was considered a big advantage of the \ac{stpa}, since it allows for the choice of a level of abstraction that is valid for every directly controlled teleoperated road vehicle. The findings of the analysis can then be integrated into a more detailed control structure to repeat the analysis. Due to the top-down approach of \ac{stpa}, further details can be integrated into the control structure later in the development process if further implementation details are known. An approach on how to iteratively perform the \ac{stpa} is presented by \citet{Thomas2015a}. Another simplification that was used throughout the analysis is in the process model variables that make up the contexts for the analysis. Using more detailed process model variables and values consequently makes it more difficult to ensure completeness, especially in urban environments, and increases the time exposure of the analysis. Nevertheless, those variables are considered very important for the outcome of the \ac{stpa}.
In spite of all these simplifications, a large amount of information is generated by performing an \ac{stpa}, considering that only the \acp{uca} resulting from one operator \ac{ca} and the \acp{cf} of one \ac{uca} were presented. Handling this amount of information becomes even more important when working with more detailed systems. The main reason for using \ac{stpa} was to include the human operator's potential flaws and errors in the analysis. Multiple \acp{uca} were identified originating in the operator not responding appropriately or timely. The analysis showed how operator-related causes such as missing/wrong training, distraction, inadequate mental model or reaction time can be determined. The \ac{stpa} also reveals how other influences such as vehicle feedback could lead to the operator reacting inadequately.

\vspace{-0.5cm}
\section*{\uppercase{Contributions}}
\noindent Simon Hoffmann initiated the idea for this paper and analyzed the teleoperated driving system. Dr. Frank Diermeyer contributed to the concept of the research project. He revised the manuscript, critically reviewing it in terms of important intellectual content, and gave final approval of the version to be published. He agrees with all aspects of the work. Artem Bykanov supported, during his master's thesis, with the execution of the STPA and upcoming discussions. We acknowledge the financial support for the project by the Federal Ministry of Education and Research of Germany (BMBF).

\bibliographystyle{apalike}
{\small
\bibliography{VEHITS2021}}


\end{document}